\documentclass{article}
\usepackage[utf8]{inputenc}
\usepackage{etex}

\pdfoutput=1
\usepackage[english]{babel}

\usepackage{amsthm}
\usepackage{hyperref}
\usepackage{color}
\usepackage{mathtools}
\usepackage{amssymb}

\usepackage{amsfonts}
\usepackage{authblk}

\usepackage{physics}
\usepackage{float}
\usepackage{enumitem}
\setlist{noitemsep} 
\usepackage{algpseudocodex}
\algrenewcommand\algorithmicprocedure{$\triangleright$}

\usepackage{graphicx}
\usepackage{subcaption}
\usepackage{bm} 
\usepackage{adjustbox}
\usepackage{multirow}

\usepackage [
  n,
  advantage,
  operators,
  sets,
  adversary,
  landau,
  probability, 
  notions,
  logic,
  ff,
  mm,
  primitives,
  events,
  complexity,
  asymptotics,
  keys
  ] {cryptocode}

\usepackage[strings]{underscore}
\usepackage{bm}
\usepackage{booktabs} 

\newcommand{\Abort}{\mathsf{Abort}}
\newcommand{\Acc}{\mathsf{Accept}}

\newcommand{\X}{\mathsf{X}}
\newcommand{\Y}{\mathsf{Y}}
\newcommand{\Z}{\mathsf{Z}}

\newcommand{\CZ}{\mathsf{CZ}}
\newcommand{\D}{\mathsf{D}}

\newcommand{\cptp}[1]{\mathsf{#1}} 

\usepackage[capitalise]{cleveref}

\usepackage{thmtools}
\declaretheorem[name=Protocol, refname={Protocol, Protocols}]{protocol}
\declaretheorem[name=Resource, refname={Resource, Resources}]{resource}
\declaretheorem[name=Routine, refname={Routine, Routines}]{routine}

\usepackage[a4paper,lmargin=1.25in, rmargin=1.25in]{geometry}

\usepackage{todonotes}
\todostyle{holl}{color=green!30, size=\footnotesize}
\todostyle{as}{color=blue!30, size=\footnotesize}

\theoremstyle{plain}
\newtheorem{theorem}{Theorem}[section]

\theoremstyle{definition}

\author{Amit Saha\thanks{Corresponding author: \texttt{abamitsaha@gmail.com}}, Harold Ollivier \\ DI-ENS, Ecole Normale Supérieure, 45 rue  d'Ulm, 75005 Paris\\ Universit\'e PSL, CNRS, INRIA}
\date{}
\title{Noise Inference by Recycling Test Rounds in Verification Protocols}

\begin{document}

\maketitle

\begin{abstract}
  Interactive verification protocols for quantum computations allow to build trust between a client and a service provider, ensuring the former that the instructed computation was carried out faithfully. They come in two variants, one without quantum communication that requires large overhead on the server side to coherently implement quantum-resistant cryptographic primitives, and one with quantum communication but with repetition as the only overhead on the service provider's side. Given the limited number of available qubits on current machines, only quantum communication-based protocols have yielded proof of concepts.
  
  In this work, we show that the repetition overhead of protocols with quantum communication can be further mitigated if one examines the task of operating a quantum machine from the service provider's point of view. Indeed, we show that the test rounds data, whose collection is necessary to provide security, can indeed be recycled to perform continuous monitoring of noise model parameters for the service provider. This exemplifies the versatility of these protocols, whose template can serve multiple purposes and increases the interest in considering their early integration into development roadmaps of quantum machines.
\end{abstract}

\section{Introduction}
\subsection{Motivation}

As quantum devices continue to advance, clients with limited quantum capabilities are increasingly dependent on remote, potentially untrusted servers to perform complex computations. At the same time, because quantum computing holds the promise of solving problems beyond the reach of classical machines, it is therefore crucial to ensure the correctness of computations delegated to these servers. However, this task becomes particularly challenging when the problem lies outside the class $\mathsf{NP}$, yet is efficiently solvable by a quantum machine, in such a case clients cannot cross-check the results using classical computers.

This challenge has been explored from a theoretical perspective, notably through the introduction of Quantum Prover Interactive Proof systems~\cite{ABE10interactive, ABEM17interactive}. Since then, various protocols have been proposed to enable the verification of quantum computations, particularly those solvable in the class $\mathsf{BQP}$, via interactive means. Some approaches, such as those in~\cite{ABE10interactive, ABEM17interactive}, rely on quantum authentication schemes, while others~\cite{FK17unconditionally} leverage Measurement-Based Quantum Computing (MBQC) to design protocols that allow clients to probe the honesty of the server. Both approaches offer information-theoretic security, though they necessitate quantum communication. A different approach was proposed in~\cite{M18classical} which relaxes this requirement, but the provided security relies on computational hardness assumptions.

Recent protocols~\cite{B18how, broadbent2025noiserobustnessdelegatedquantumcomputation, LMKO21verifying, KKLM22unifying} have advanced the former approach, incorporating multiple rounds of interaction, some dedicated to computation and others to testing the server's behavior. This paradigm, termed \emph{Compute and Test}, is particularly suited for near-term quantum devices. It avoids the space overhead associated with earlier approaches by shifting it to repeating computation and testing rounds. However, even with a repetition count growing as $O(\log(1/\epsilon))$, where $\epsilon$ is the distinguishing probability between the ideal device and the real protocol, the burden of repeated rounds remains significant, and further reductions are desirable.

From the server’s perspective, verification protocols not only enforce honesty but also demand precise quantum control to ensure that instructions are executed faithfully. Any failure to do so could raise suspicions about the correctness of the computation. Practically, this necessitates constant monitoring and calibration. Since calibration processes are typically separate from the computation itself, servers must periodically go offline to perform these checks, introducing additional overhead. 

The central question we address in this paper is whether it is possible to reduce this offline time by using test round data, collected by the client, to assist in the server's calibration process. We demonstrate that this is indeed possible by building on the idea presented in~\cite{KKLM22unifying}, in the \emph{Compute and Test} paradigm, verification relies on error detection, which can be leveraged to infer errors. In our case, this enables device characterization, providing a pathway to reduce calibration downtime while maintaining protocol security.

\subsection{Related Work}

\paragraph{Verification.}
This work relies on using the protocol introduced in~\cite{LMKO21verifying} as \emph{robust Verifiable Blind Quantum Computing} (rVBQC) \cite{BFK09universal}. It follows the Compute and Test paradigm by randomly interleaving computation and test rounds while delegating their execution to the server through \emph{Universal Blind Quantum Computing} (UBQC). This guarantees that both types of rounds are indistinguishable from one another.

More precisely, in the simplest scenario, the client will run the protocol using $N$ rounds, using randomly half of them for the computation and the other half for the tests. The candidate result of the computation is obtained by aggregating the outcomes of all the computation rounds and taking the majority vote. One can then remark that because $\mathsf{BQP}$ instances have a correctness-soundness gap, to corrupt the candidate result, the server must attack successfully and cause a bit-flip for a number of computation rounds that grows linearly in $N$.

Regarding the tests, they are chosen in such a way that they correspond to Clifford computations whose execution yields a deterministic outcome for some final measurement. Because they are run encoded, the server does not know the value of the outcome that is expected by the client, allowing the latter to check the honesty of the former. Indeed, whenever the received outcome does not correspond to the expected one, the client can be sure that the server has not been implementing the given instructions faithfully, thereby detecting a possible attack.

The security proof then works by assessing how likely it is to attack enough computation rounds to cause a flip of the result while being undetected. Equivalently, it determines the probability of successfully flipping the result of the candidate computation when fewer than a given maximum tolerated number of test rounds failed. Computing these probabilities explicitly shows that whenever the maximum tolerated number of failed test rounds is below a threshold value that depends on the correctness-soundness gap, this probability becomes exponentially small in $N$. As a consequence, it is safe to accept the candidate result when the observed number of failed test rounds is below the maximum tolerated value.

This protocol and its security guarantees are detailed in \cref{sec:verification}. Compared to other works such as~\cite{ABE10interactive, ABEM17interactive, B18how} or those stemming directly from~\cite{FK17unconditionally}, such as~\cite{KW17optimised}, it combines several appealing properties:
\begin{enumerate}
\item It is composable, meaning that it can be used in a wider context without having to reprove its security in this wider context;
\item It is efficient as the overhead grows only as \(\log(1/\epsilon)\) for \(\epsilon\) the desired security error;
\item It is scalable in the sense that the overhead measured by the number of rounds is independent of the size of the computation; and
\item It is robust meaning that as long as the noise is gentle enough so as not to cause too many failed test rounds, the protocol will accept with high probability. 
\end{enumerate}   

\paragraph{Noise Characterization.}
Characterization of quantum devices is an essential task in the process of building quantum machines and very large amount of techniques have been developed to tackle various flavors of the initial question \cite{Wallman2016NoiseTailoringRandomizedCompiling}.

Current techniques can be cast into two broad families. One that works at the component level, while the second works at the system level. More concretely, component-level techniques study a small functionality in isolation. The idea it to fully determine the behavior of the device to improve its behavior. Because the components are generally of modest size, the efficiency of the technique is not too much of a concern. The main limitation of this approach is that it does not give guarantees about the behavior of the component when used in conjunction with other components. For instance, component-level characterization of a 2-qubit gate would not be able to fully anticipate effects such as cross-talk that are inherent to the combination of many components close to one another.

The second family has been precisely developed to characterize the behavior of the full system. There, one will want to keep the arrangement of the components as they are when the system is used, while ensuring some degree of characterization of its behavior. The information extraction task is usually more complex because efficiency becomes a major concern. This is in part because writing a generic process as a CPTP map would require determining an exponential number of a priori independent coefficients. Indeed, system-level techniques overcome this difficulty by assuming that the process has a simpler form---e.g., it is a Pauli channel \cite{Fawzi2023LowerBoundsPauliChannels, Flammia2020EfficientEstimationPauliChannels, Chen2022LearnabilityPauliNoise, Flammia2021PauliErrorEstimationPopulationRecovery, Harper2020EfficientLearningQuantumNoise, Harper2019EfficientLearningGeneralizedPauliChannels, Harper2023LearningCorrelatedNoise39Qubit, Wagner2023LocalityStructuredPauliNoise} and possibly making further assumptions such as Markovianity \cite{rouze2023efficientlearningstructureparameters}.

In this paper, we will be interested in system-level characterization. Randomized benchmarking and its variants have been used to great success in many concrete situations \cite{Franca2018ApproximationRandomizedBenchmarking, Helsen2019SpectralRB, Helsen2022GeneralFrameworkRB, Magesan2012EfficientMeasurementGateError, Heinrich2023RandomizedBenchmarkingRandomQuantumCircuits}. Other approaches include Average Circuit Eigenvalue Sampling (ACES) \cite{Flammia2021AverageCircuitEigenvalueSampling}. This method works specifically for Clifford circuits with Pauli noise. It leverages the following:
\begin{enumerate}
\item A single Pauli $\cptp{P}$ operator is transformed by the Clifford operation $\cptp{C}$ into an another single Pauli operator $\cptp{P}'=\cptp{C}[\cptp{P}]$ giving rise to a generalized eigenvalue equation---here associated to eigenvalue 1;
\item A Pauli operator $\cptp{P}$ is transformed under a Pauli noise channel $\cptp{N}$ into $\cptp{N}[\cptp{P}] = \lambda(\cptp{P}) \cptp{P}$ with $\lambda(\cptp{P}) \in [-1;+1]$;
\item Generic noise can be transformed into Pauli noise through twirling.
\end{enumerate}
Combining these elements, the effect of a noisy Clifford circuit $C$ implementing \(\cptp{C}\) through \(n\) layers \({\cptp{C}}_i\) such that $\cptp{C} = \cptp{C}_n \cptp{C}_{n-1} \ldots \cptp{C}_2 \cptp{C}_1$ is to transform $\cptp{P}$ into
\begin{equation}
  \Lambda_{C}(\cptp P) \cptp{C}[\cptp{P}] =
  \lambda_{C_n}(\cptp{C}_{n-1} \ldots \cptp{C}_1[\cptp{P}])
  \lambda_{C_{n-1}}(\cptp{C}_{n-2} \ldots \cptp{C}_1[\cptp{P}])  \ldots
  \lambda_{C_2}(\cptp{C}_1[\cptp{P}])
  \lambda_{C_1}(\cptp{P}) \cptp{C}[\cptp{P}].\label{eq:aces}
\end{equation}

With this at hand, the characterization procedure proposed by ACES for a given $\lambda_{\cptp{C}_i}$ works by
\begin{enumerate}
\item Constructing circuits comparable to the target circuits where one wants to use the corresponding gate;
\item Evaluating through observable estimation the value of \(\Lambda_{C}(\cptp{P})\) for various $\cptp{C}$ and $\cptp{P}$ that involve a contribution of \(\lambda_{C_i}\);
\item Numerically inverting the system of equations to recover an estimate of $\lambda_{C_i}$. 
\end{enumerate}
The original paper \cite{Flammia2021AverageCircuitEigenvalueSampling} proposes to solve such a system of equations for several values of $\lambda_{C_i}$ at once by taking their logarithm, which effectively transforms it into a linear system of equations whose coefficients form the \emph{design-matrix}. 

\subsection{Contributions}
This paper addresses the questions of what information can be extracted from verification protocols about the behavior of the server and how it can be used in practical situations. The interest of this question comes from the apparent gap between what a verification protocol provides, whether the computed result is correct, or the protocol should abort and how verification protocols are designed in the Compute and Test paradigm through the use of error detection codes~\cite{KKLM22unifying}. In essence, verification protocols only output whether the server has been acting honestly or maliciously, while it uses tools that are usually used to infer detailed information about errors and can participate in their correction.

\paragraph{Verification protocols can learn hardware defects once given a possible noise model.}
It is recognized that the origin of the informational gap lies in the discrepancy between what the client and server are willing to believe. Indeed, the client sees the server as arbitrary and malicious, which is deeply incompatible with extracting a large amount of information. In other words, it is not meaningful to ask about the information gained by the client about the server beyond whether it is honest or not. On the contrary, the server knows whether it is actively malicious or passively noisy. In this latter case, it is meaningful to ask whether the server can learn the parameters of the noise model that it experiences. This is precisely the scenario that we explore in the next sections. A suspicious client that wants to keep being protected against arbitrary deviations from the server, and an honest but noisy server that wants to extract as much information as possible from the verification procedure to update its noise model.

\paragraph{Noise model parameter learning can be made efficient for a large class of noise model and computation graphs.}
As the setting is now established on firm ground, the question is how to proceed with the information extraction. It is shown in the next section that ACES can be embedded into the verification protocol at no cost. More precisely, a slight change in the verification protocol will modify the execution of the test rounds in a way that keeps their properties intact for verification by the client while also being interpretable as ACES rounds, collecting data to form the design-matrix.

As a result, \cref{proto:vbqc-aces} has the following properties:
\begin{enumerate}
\item It constructs Secure Delegated Quantum Computation (\cref{res:sdqc}) with an error negligible in $N$, the number of rounds of the protocol, meaning that it is secure against an arbitrary malicious server;
\item It is robust to constant circuit-level noise;
\item It allows an honest-but-noisy server to update parameters of its noise model.
\end{enumerate}

In practice, \cref{proto:vbqc-aces}, while performing standard verification for the client, informs the server about calibration drifts at no extra cost, thereby reducing the downtime necessary to perform recalibration.

\section{Integration of ACES into rVBQC}
In this section, we show how ACES can be integrated into rVBQC. To this end, we will proceed in 3 steps: first, we review briefly rVBQC and its security guarantees; second, we show that some information about noise can be extracted by the protocol from the point of view of the server in a way that resembles ACES; third we show that, while the structural rigidity of rVBQC seems to jeopardize ACES, it is indeed possible to blend the two protocols into a new one that:
\begin{enumerate}
\item Achieves the same security and noise-robustness guarantees as the original rVBQC;
\item Allows the server to extract noise parameters for a wide variety of Pauli channels.
\end{enumerate}

\subsection{Verification and rVBQC}
\label{sec:verification}

The security guarantees of rVBQC are given in the Abstract Cryptography (AC) framework \cite{MR11abstract-cryptography}. This framework not only provides a general structure for security proofs, but most interestingly shows that whenever a protocol is secure in AC, then it is inherently composable. More concretely, this means that this protocol can be safely used within the context of a larger protocol without having to reprove its security in this new context. This contrasts with stand-alone security proofs whose security guarantees hold only in isolation.

The high-level methodology imposed by AC is to first define an ideal resource. It is nothing more than a mathematical specification of what the final device is supposed to do. For instance, when a client asks for verifying a computation \(\cptp{C}\) delegated to a server, this means that either it receives the correct result or it receives \(\Abort\). This means that it is never forced to accept an incorrect result. This is also the best that it can ask, as it cannot force the server to perform the computation, e.g., the server can always refuse to send the computation result. Additionally, rVBQC provides blindness, which means that the server ``does not know'' the computation \(C\) being delegated. Obviously, the server will still at least have an upper bound on the size of the computation simply because it can measure the time it takes per round. To encompass this unavoidable leakage of information, blindness will be relative to a set of computations \(\mathfrak{C}\) in the sense that while delegating \(\cptp{C}\in \mathfrak{C}\), the server learns at most \(\mathfrak{C}\).

All of this can now be summarized in the following Secure Delegated Quantum Computation resource:
\begin{resource}[Secure Delegated Quantum Computation]
  \label{res:sdqc}
  \item
  \begin{algorithmic}[0]
    \State \textbf{Public information}: Nature of the permitted leakage.
    \State \textbf{Permitted leakage}: Set of computations \(\mathfrak{C}\).
    \State \textbf{Client's input}: A classical description of \(\mathsf{C} \in \mathfrak{C}\).
    \State \textbf{Server's input}: Two bits \(c,d\).
    \Procedure{Resource's operations}{}
    \State It receives the client's input.
    \State If it receives \(c=1\), it sends the leakage to the server who returns \(d\).
    \State If \(d=1\), it outputs \(\Abort\) at the client's interface.
    \State If \(c=0\) or \(c=1 \wedge d=0\), it outputs the result of computation \(\mathsf{C}\) and \(\Acc\) at the client's interface.
    \EndProcedure
  \end{algorithmic}
\end{resource}
This resource is deemed \emph{secure-by-design} because it is the transcription of the input-output relationships that can be deduced from the plain-English description of the functionality it corresponds to. 

In particular, AC specifies that the definition of an ideal resource does not propose an implementation. The implementation is actually the second step in the AC methodology and corresponds to the protocol. For rVBQC, it works in the following way:
\begin{protocol}[rVBQC]
\label{proto:rvbqc}
\item
  \begin{algorithmic}[0]
    \State \textbf{Public information}: 
    \begin{itemize}
    \item A graph \(G\) and a flow \(f\).
    \item Integers \(N,d,w\) with \(d < N\) and \(w  < N-d\) representing the total number of rounds, the number of computation rounds, and the number of tolerated failed test rounds.
    \end{itemize}
    \State \textbf{Client's input}: \(P\), a MBQC pattern over \(G\) and compatible with the flow \(f\).
    \State The client samples the location of \(d\) computation rounds in \([N]\).
    \Procedure{Round Delegation}{}
    \State If the round is a computation round, prepare all qubits in \(\ket+\) and delegate the MBQC pattern \(P\) to the server using UBQC (\cref{proto:ubqc}).
    \State If the round is a test round, perform the qubit preparation and obtain the MBQC pattern \(T\) using \cref{routine:test-sampling}, then delegate it to the server using UBQC (\cref{proto:ubqc}), both detailed in \cref{sec:ubqc}.
    \EndProcedure
    \Procedure{Deviation detection}{}
    \State The client counts the number of failed test rounds (i.e., rounds for which at least one of the trap measurements does not correspond to a +1 outcome).
  \State If the number of failed test rounds is greater than \(w\), output \(\Abort\).
  \EndProcedure
  \Procedure{Majority vote}{}
  \State If the protocol has not aborted, compute the majority vote of the results of computation rounds and output it together with \(\Acc\).
  \EndProcedure
  \end{algorithmic}
\end{protocol}
First, note that the set of computations  \(\mathfrak{C}\) for which the protocol is blind is relative to all computations that can be performed using a graph state \(\ket G\) and a fixed flow $f$. In essence, once the security proof is provided, this means that the protocol will leak at most \(\ket G\) and $f$ but not the specific values of the measurement angles. Because there are universal graph states that have a fixed structure and can be enlarged to compute any unitary, this amounts to leaking an upper bound on the size of the computation.

Going through the various steps, the protocol starts by selecting computation rounds and test rounds at random among the total \(N\) rounds of the protocol. For the test rounds, an additional choice is made as there are a total of \(\chi\) types of test rounds, where \(\chi\) is the chromatic number of \(G\). The purpose of these test rounds is to be a collection of several traps, each of them corresponding to a very simple computation, namely, the identity for the trap qubit initially prepared in \(\ket +\), while being sensitive to deviations of the server. Each corresponding MBQC pattern is then delegated to the server using a secondary protocol, Universal Blind Quantum Computing (UBQC), see \cref{sec:ubqc}, whose sole purpose is to provide indistinguishability between these rounds from the server's point of view. The protocol will return \(\Abort\) whenever the number of failed test rounds exceeds some threshold value \(w\) that is determined as a function of \(\chi\) and the completeness-soundness gap \(c\) of the problem that \(\cptp{C}\) solves. Otherwise, it returns the majority vote of the computation rounds.

The intuition behind the security of rVBQC is the following. First, the existence of a constant sized completeness-soundness gap ensures that to flip the result of the majority vote, the server will need to flip many of the individual computation rounds. But because of blindness, to achieve this requires attacking a large number of rounds, be they computations or tests. As tests are designed to be sensitive to deviations, they will detect the malicious server.

The actual full proof is structured around the concept of simulation; this is the heart of the AC framework. Indeed, what a client would like from a secure protocol is that it implements the corresponding ideal resource, in the sense that both are equivalent from the perspective of distinguishing the specification from its implementation. This makes sense because the resource is secure by design, so that if a protocol is indistinguishable from the ideal resource, then it will inherit the security of the resource. In practice, this amounts to finding a simulator that can be attached to the interface of the ideal resource facing the server so that this simulator generates a transcript that is indistinguishable from what a malicious party would see when interacting with the protocol. If this is the case, it is safe to assume that the protocol \emph{constructs} the ideal resource. The same definition holds when the transcript's indistinguishability is replaced with diamond-distance bounded by some \(\epsilon\).

For the case of rVBQC, we have:
\begin{theorem}[Security of rVBQC~\cite{LMKO21verifying}]
rVBQC constructs SDQC with error \(\epsilon \in \negl[N]\).
\end{theorem}

\subsection{Extracting information about the noise}

\paragraph{Effective noise in rVBQC.}
A crucial characteristic of rVBQC is that all quantum states are one-time-padded in the sense that states at any point during the computation bear a qubit-by-qubit encryption. A qubit in state \(\rho\) is ciphered into $\Z^z \X^x \rho \X^x \Z^z$ where \(x,z\) are bits that are sampled uniformly at random for each qubit. The impact on the noise perceived \emph{after} removing the encryption is to twirl the noise, transforming it into a convex combination of Pauli-strings applied right before the measurement in the $X$ basis. Combined with randomized compiling for the gates creating the graph state, it is enough to restrict the study to Pauli noise models, as randomized compiling ensures the noise is not only twirled before the measurement, but also \emph{between the gates}.

\begin{figure}[!h]
	\centering
	\includegraphics[scale=.32]{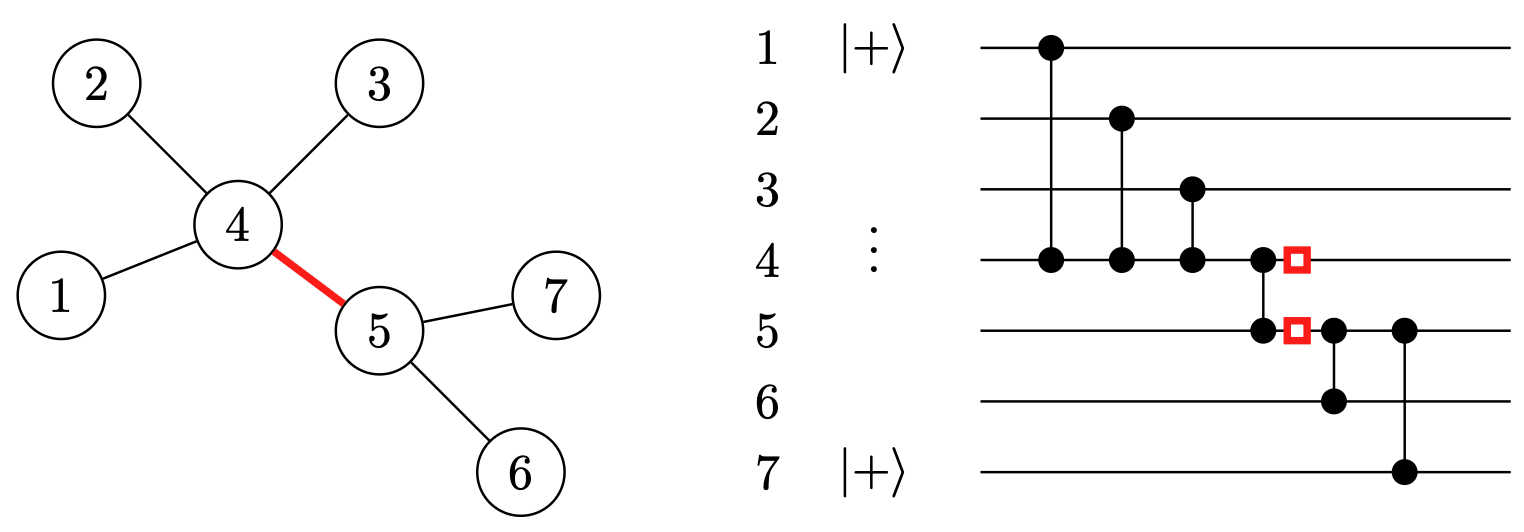}
\caption{An example of a single faulty entangling gate in graph state preparation. A graph state $|G\rangle$ is prepared by applying CZ gates to qubits initialized in $|+\rangle$. All CZ gates are assumed ideal except one edge $f=(4, 5)$, whose implementation is modeled as a perfect $\mathrm{CZ}_{(4,5)}$ followed by local depolarizing channels on the qubits it touches.}
	\label{fig:noisy-gate}
\end{figure}

\paragraph{Foundational example.}
To build intuition, consider the situation where a single faulty $\CZ$ gate appears in the construction of a large graph state $\ket G$. More precisely, for a graph  \(G = (V,E)\) and an edge \(e  \in E\), \(\CZ_e\) is the controlled-\(\Z\) gate acting on the qubits \((v,w)=e\). Suppose that for all but one \(e\), the applied gate is perfect, while for some \(f=(r,s) \in E\) the applied gate is modeled by a perfect \(\CZ\) gate followed by two depolarizing channels \(\cptp{D}\) with strength \(\lambda_r\)  and \(\lambda_s\) respectively, i.e., instead of applying \(\CZ_{f}\) we have \((\D_{\lambda_r} \otimes \D_{\lambda_s}) \circ \CZ_f\).This situation is depicted in \cref{fig:noisy-gate} for a small 7-qubit graph state.

\begin{figure}[H]
	\centering
    \includegraphics[scale=0.32]{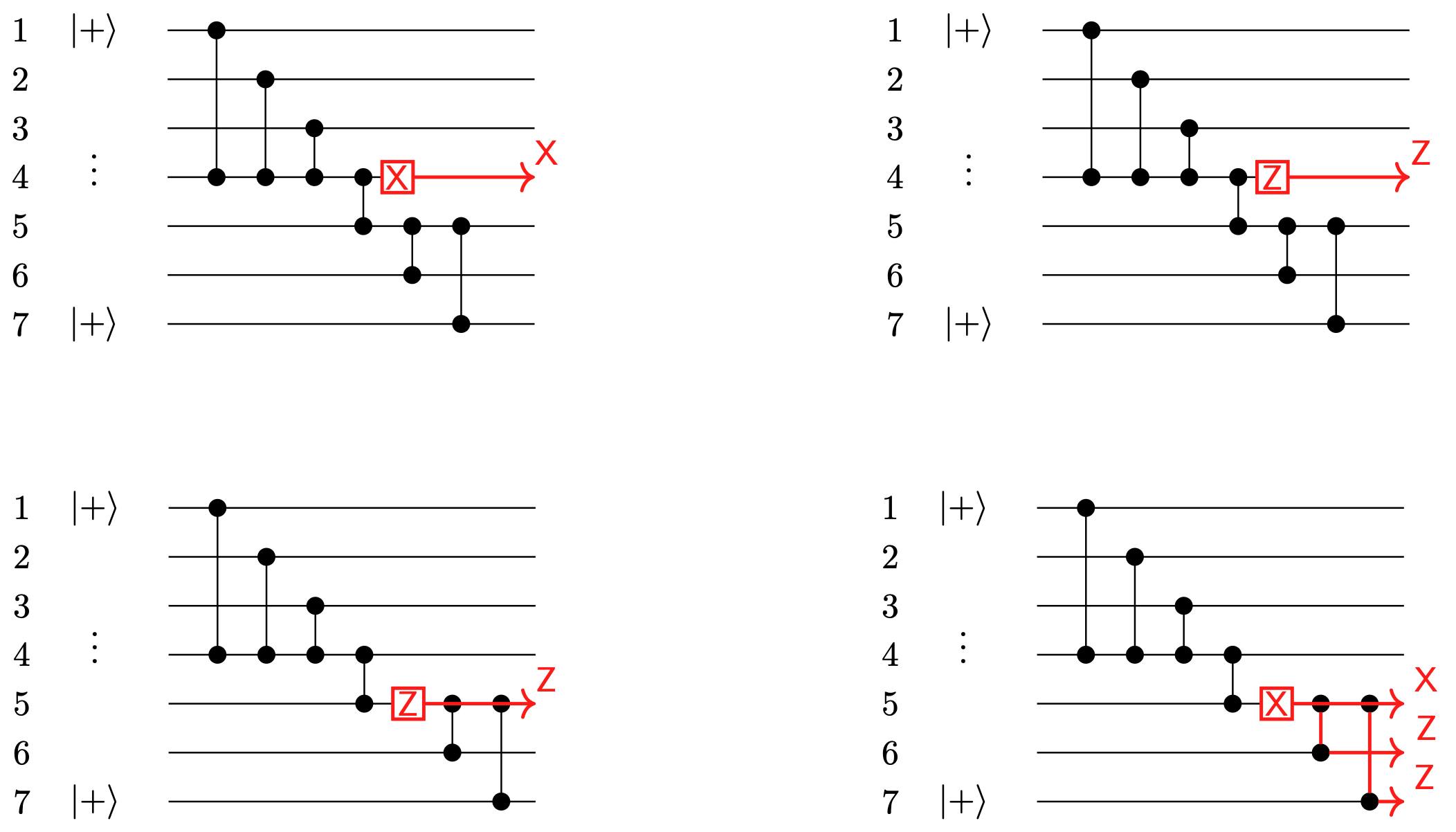}
\caption{Propagation of Pauli errors after the faulty $\CZ_{(4,5)}$ gate. The red boxed Pauli operator indicates the error immediately after the faulty gate, and each panel shows its propagation through the remaining $\CZ$ gates in the circuit. A $\X$ error on qubit $4$ (Top left). A $\Z$ error on qubit $4$ which remains localized on $4$ (Top right). A $\Z$ error on qubit $5$ which remains localized on $5$ (Bottom left). An $\X$ error on qubit $5$, which propagates forward through subsequent $\CZ$ gates,
	leaving $\Z$ errors on later neighboring qubits $6$ and $7$ while the $\X$ error remains on $5$ (Bottom right).}
	\label{fig:noise-propagation} 
\end{figure}

Using the commutation relations of the \(\CZ\) gates with individual Pauli unitaries, it is easy to derive the impact of the noise happening on each qubit of a test round, recalling that all non-encoded measurement angles are then equal to 0. A \(\Z\) directly flips the measurement outcome in the \(\{\ket +, \ket - \}\) basis, while a \(\X\) has no effect on the measurement on the affect qubit, but propagates to neighboring qubits as further \(\CZ\) gates are encountered in the rest of the circuit creating \(\ket G\). This is illustrated in \cref{fig:noise-propagation}.

As a consequence, it becomes easy to identify a faulty gate by observing in rVBQC which traps are failed and which ones are not. In fact, one can notice that some traps probability are directly linked to \(\lambda_r\) and \(\lambda_s\). Using the example in \cref{fig:noisy-gate}, and the propagation of the Pauli unitaries, a bit flip for the trap in \(s\) can only happen when the depolarizing channel on qubit \(s\) yields a \(\Y\) or \(\Z\). This happens with probability \((1-\lambda_s)/2\), so that observing the failure probability of the trap in \(s\) is a direct observation of the noise happening on \(\CZ_{(r,s)}\).

Yet, for this to be possible, some information has to be communicated back to the server, as it has no access to the unencrypted data that is necessary to declare if a trap has failed or not. One could imagine that the client is well placed to perform the evaluation of the trap failure probabilities, as it is already keeping test round failures. Instead, the client could simply communicate the raw data of the test rounds, i.e., the randomness used to create the rounds comprising the encryption keys to the server and let it decrypt its measurement outcomes and estimate the failure probability directly. The important remark is that this does not affect the security of the whole scheme. As long as this transmission happens after the completion of rVBQC, the computation rounds are uncorrelated to the test rounds, and any post-processing from the server cannot create additional correlations to the computation rounds. Hence, the security is not affected by the release of the keys used for the test rounds after the protocol is completed.

This foundational example revealed two important aspects of verification:
\begin{enumerate}
\item Trap failure rate reveals information about the noise that affects the gates used to create and compute with the graph state;
\item Trap failure rate is sensitive to the order in which the noisy gates are applied as \(\X\)-noise only propagates forward in the circuit to qubits that are direct \emph{future} neighbors of the noisy one.
\end{enumerate}

\subsection{Performing ACES within rVBQC}
The previous subsection hinted at the possibility of extracting noise information from trap rounds. Yet, in order to materialize this into a concrete protocol, two gaps need to be addressed:
\begin{enumerate}
\item The relationship between the failure probability of a given trap and the noise parameters;
\item The generation of an invertible design matrix to recover individual noise parameters;
\end{enumerate}
while keeping the security of the whole protocol.

\paragraph{Trap failure probability.}
The traps in rVBQC correspond to preparations that ensure a deterministic result for an observable being measured by the server. For instance, in the simplest case, a trap at qubit \(v\) means that it is expected that for well prepared qubits, an honest server should deterministically obtain the +1 outcome when measuring the observable \(X_{v}\).\footnote{Recall that this description corresponds to the instructions that would indeed be delegated, and thus encrypted, using UBQC.} In other terms, this means that the input qubits state \(\ket{\psi(v)}\) for generating a trap at qubit \(v\) needs to satisfy:
\begin{equation}
\X\left[\cptp{C} [\ketbra{\psi(v)}] \right] = \cptp{C}[\ketbra{\psi(v)}],
\end{equation}
where \(\cptp{C} = \prod_{e\in E,\prec}\CZ_{e}\), and \(\prec\) is an ordering of $E$.

As noted in \cref{sec:test-rounds}, if one prepares the product state \(\ket{+}_v \bigotimes_{(v,w)\in E} \ket{0}_w\) as input and feeds it through the perfect circuit that generates \(\ket{G}\), it is unchanged. Now, recall that if a state is stabilized by an abelian subgroup \(S\) of the Pauli group, the projector onto that state can be written as
\begin{equation}
\frac{1}{|S|}\sum_{\cptp P\in S} \cptp P.
\end{equation}
This representation makes it straightforward to infer the output when noise affects precisely those gates that act non-trivially on the relevant stabilizers. Indeed, fix a stabilizer element \(\cptp P\in S\) and let \(C\) denote the sequence of \(\CZ\) gates chosen by the server to implement the graph state. After the noisy implementation, the contribution associated with \(\cptp P\) becomes \(\Lambda_{C}(\cptp P) \cptp P'\), where \(\cptp P'\) denotes the image of \(\cptp P\) under the ideal Clifford circuit that prepares \(\ket{G}\). It follows that the output state produced by the noisy circuit for the input state preparing a trap at \(v\) can be expressed as
\begin{equation}
\rho_{\mathrm{out}}(v) = \frac{1}{|S|} \sum_{\cptp P\in S}\Lambda_C(\cptp P)\cptp P'.
\end{equation}
Therefore, the trap failure probability satisfies
\begin{align}\label{trap_failure}
  p_{\mathrm{fail}}(v)
  & = \frac{1}{2|S|} \tr\left((\cptp I - \cptp X_v) \rho_{\mathrm{out}}(v) \right) \\
  & =\frac{1-\Lambda_{C}\left(\X_v\bigotimes_{(v,w) \in E}\Z_w\right)}{2},
\end{align}
where \(p_{\mathrm{fail}}(v)\) denotes the probability that the trap at \(v\) fails, and we used the facts that \(\tr(\cptp P)=0\) for all \(\cptp P \neq \cptp I\), while the identity contribution yields \(|S|\). In particular, the bias $P_{v}(C)$ between trap success and failure is directly given by
\begin{equation}
P_v(C) = p_{\mathrm{succ}}(v)-p_{\mathrm{fail}}(v) = \Lambda_{C} \left( \X_v \bigotimes_{(v,w) \in E} \Z_w \right).
\end{equation}	
Since traps correspond to stabilizers of the graph state, their evolution through the preparation circuit is governed by the usual conjugation rules for \(\CZ\) gates. The only degree of freedom available to the server that remains compatible with blindness is the ordering of commuting \(\CZ\) gates. Distinct orderings generally produce distinct linear constraints. Hence, by varying these orderings sufficiently, one can obtain a family of linearly independent equations that is rich enough to estimate all noise parameters \(\lambda\), provided that no intrinsic symmetry of the construction enforces unavoidable degeneracies.



\paragraph{Generating an invertible design matrix.}
While the previous paragraph demonstrated that rVBQC allows to perform one crucial step of ACES evaluation of some \(\Lambda\)'s, to complete the extraction of noise model parameters, it is necessary to invert the relationship between the gate-level noise model, i.e.,~\(\lambda_{e}\), and the circuit-level noise model, i.e.,~\(\Lambda\), see \cref{eq:aces}. When ACES is applied in a stand-alone fashion, this is usually done by varying the circuit applied and the input states. Yet, in the scenario described here, there is no such freedom. The reason is that the circuit \emph{must} implement the Clifford \(\cptp{C}\) that creates the graph state \(\ket G\) from \(\ket{+}^{|V|}\) states. The reason is that rVBQC imposes the structure of test rounds containing the traps. Adding extra rounds designed solely for noise estimation purposes is always possible, but they wouldn't correspond to recycling data from rVBQC itself.

Hence, the question of generating an invertible generating matrix should be considered with the additional constraint imposed by rVBQC, that is, with a fixed target Clifford \(\cptp{C}\) and using \(\X\) measurements only, so that these correspond to data generated by rVBQC traps.  Recalling \cref{eq:aces} and the example of \cref{fig:noisy-gate} and \cref{fig:noise-propagation}, we see that one possibility is to change the order of the \(\CZ\) gates that implement \(\cptp{C}\). While for any permutation \(\sigma: E \rightarrow E\) we have,

\begin{equation}
\cptp{C} = \prod_{e \in E, \prec} \CZ_e = \prod_{e \in E, \prec} \CZ_{\sigma(e)}, 
\end{equation}
where \(\prec\) is an order over \(E\), their corresponding noisy versions, respectively defined as \(\tilde{\cptp{C}}\) and \(\tilde{\cptp{C}}_{\sigma}\) do not match. This is because, as noted before, noise propagates to future neighbors, which depends on the order in which \(\CZ\) gates are applied. By choosing various permutations \(\sigma \in \Omega\), where \(\Omega\) is a subset of the permutations of \(|E|\), it then becomes possible to generate a different equation linking the observed probabilities of success and failures of a trap at a given qubit \(v\) given the chosen \(\sigma\). Using this freedom, ACES is integrated into rVBQC in the following way:
\begin{protocol}[rVBQC with ACES]
  \label{proto:vbqc-aces}
\item
  \begin{algorithmic}[0]
    \Procedure{Server Initialization}{}
    \State The server chooses randomly a permutation \(\sigma_i \in \Omega\) for each \(i \in [N]\).
    \EndProcedure
    
    \Procedure{rVBQC Execution}{}
    \State The Client and the server perform rVBQC, where the server applies the \(\CZ\) gates in the order defined by \(\sigma_i\) for round \(i\).
    \EndProcedure
    
    \Procedure{Randomness Sharing}{}
    \State The client sends all the randomness used during the test rounds to the server.
    \EndProcedure
    
    \Procedure{ACES}{}
    \State The server computes the success and failure probabilities for each trap \(v\) and each order \(\sigma\), and inverts the corresponding design matrix to recover the gate-level noise parameters.
    \EndProcedure
  \end{algorithmic}
\end{protocol}

Note that in addition to the order of \(\CZ\) gates, one could perform other changes that still implement the same unitary \(\cptp{C}\), while increasing the capability to generate more equations linking observed success and failure probabilities for traps and gate-level noise parameters. For instance, one could replace any \(\CZ\) gate by an odd number of the same \(\CZ\) gate, and apply shifts on top of this. We will comment on this ability at the end of the next section. This remark nonetheless outlines the important remaining question, whether one can estimate noise parameters for a reasonably complex model in spite of the very restricted sets of circuits and measurements that are compatible with the security of rVBQC. We will answer this question by instantiating rVBQC with ACES for a specific noise model in the next section. The reason for not seeking a theoretical answer for it is that for ACES alone, i.e.,~without rVBQC, the answer will indeed be strongly dependent on the noise model. As the number of parameters to estimate increases, the task becomes more complex, and more equations need to be generated. For purposefully complex noise models, ACES can become intractable, while for reasonable noise models, it is generally not the case.

\section{Learning Gate Level Depolarizing Noise}\label{sec:Depnoise}

\paragraph{Noise model.}
In this section, we explore a model where \(\CZ\) gate implementation is noisy. More precisely for \(e = (v,w) \in E\), the noisy \(\CZ_{e}\) is obtained by applying the perfect one followed by depolarizing channels acting on a set of qubits \(\nu(e) \subset V\), the depolarizing strength being defined by \(\lambda_{e,u}\) for $u\in \nu(e)$. When \(\nu(e) = \{v,w\}\), this corresponds to applying a depolarizing channel with different strengths for each qubit touched by the gate. The situation \(\{v,w\}  \subsetneq \nu(e)\) corresponds to having cross-talk as qubits that are not involved in the gate are also affected by some noise.

\paragraph{Estimating the depolarizing strength in the absence of cross-talk.}
Let \(v \in V\) be a qubit. Following the notation used in the previous section, \(p_{\mathrm{succ}}(v)\) and \(p_{\mathrm{fail}}(v)\) are the corresponding success and failure probabilities of the trap in qubit \(v\).

First,  assume \(v\) is of degree 1. In such a case, there is a single \(w\) such that \(e =  (v,w) \in E\). To estimate \(\lambda_{e,v}\), arrange the gates so that \(\CZ_e\) is applied before any \(\CZ_f\) where \(f\) is an edge involving \(w\). Denote by $C$ the corresponding circuit. In this case, the only noise affecting the trap at \(v\) is coming from \(\CZ_e\), so that we have:
\begin{equation}
\lambda_{e,v} = p_{\mathrm{succ}}(v) - p_{\mathrm{fail}}(v) = P_v(C),
\end{equation}
where the probabilities are computed keeping only the test rounds in the prescribed order.

For \(v\) of degree strictly greater than 1, there are \(e,f \in E\) such that \(e = (u,v)\) and \(f = (v,w)\) with \(u \neq w\). Consider the case where \(\CZ_e\) is applied before \(\CZ_f\) and so that both are applied after any  other \(\CZ\) touching \(u\); this happens for instance if \(\CZ_e\) and \(\CZ_f\) are the last gates of \(C\).  The depolarizing noise acting on \(v\) after \(\CZ_{f}\) cannot propagate to \(u\) through \(e\). Conversely, when \(\CZ_f\) is applied before \(\CZ_e\) it does. Denoting by $C$ the circuit corresponding to the first case, \(C'\) the second, all other gates being applied in the same order, we conclude that:
\begin{equation}
\Lambda_{C'}(\cptp{P}_u) = \lambda_{f,v}\Lambda_{C}(\cptp{P}_u), 
\end{equation}
where \(\cptp{P}_u = \X_u \bigotimes_{(u,t)\in E} \cptp{Z}_t\) is the preparation for the trap in qubit \(u\).

As a result, by varying the order of \(\CZ\) gates we see that it is possible to estimate any of the parameters \(\lambda_{e,v}\) for \(e = (v,w) \in E\).

\paragraph{Estimating the depolarizing strength in the presence of cross-talk.}
The construction of cross-talk noise is similar to that of the previous case, taking into account the fact that \(\nu(e) \) is not restricted to the nodes of \(e\). Indeed, for \(f = (v,w)\), \(\lambda_{f,v}\) can be estimated as before. For \(u \in \nu(f)\setminus \{v,w\}\), consider an edge \(e =(t,u)\) with \(t \notin \nu(f)\). Then fix the circuits \(C\) so that \(\CZ_e\) is applied before \(\CZ_f\) and \(C'\) where \(\CZ_f\) is applied before \(\CZ_e\), everything else remaining identical, including the fact that these gates are applied after all \(\CZ\) gates that affect \(t\), one can deduce that:
\begin{equation}
\Lambda_{C'}(\cptp{P}_t) = \lambda_{f,u}\Lambda_{C}(\cptp{P}_t), 
\end{equation}
where \(\cptp{P}_t = \X_t \bigotimes_{(s,t)\in E} \cptp{Z}_{s}\) is the preparation for the trap in qubit \(t\).

\paragraph{Optimizing Test Rounds and Orderings.}
The technique presented above shows that, without cross-talk, given a qubit and an edge of the graph \(G=(V,E)\) used in the computation, the corresponding depolarizing noise strength can be evaluated by changing the order of a single gate that contains that qubit with respect to the one for which the parameter is estimated. Once again, because noise only propagates to future neighbors, this requires fixing the order of all the gates that share a qubit with any of the two edges involved in the estimation.

As a consequence, the full procedure can be parallelized. Consider \(W\) a set of ordered triples \(\{(u,v,w), (u,v) \in E,  (v,w) \in E\}\) such that
\begin{equation}
  \forall v \in V, e = (v,w) \in E, \ \exists u \in V \mbox{ with } (u,v,w) \vee (w,v,u) \in W \label{eq:W-condition}.  
\end{equation}
By construction, each element \(h = (u,v,w) \in H\) allows to extract \(\lambda_{(u,v),v}\) and \(\lambda_{(v,w),v}\), so that the condition of \cref{eq:W-condition} corresponds to being able to estimate all the parameters of the gates used in the construction of the graph state \(\ket G\). Additionally, if \(h,h' \in W\) have no common vertex, the ordering can be fixed independently of one another, so that only two orderings are necessary to extract the information about all four parameters.

\begin{figure}[H]
	\centering
    \includegraphics[scale=.32]{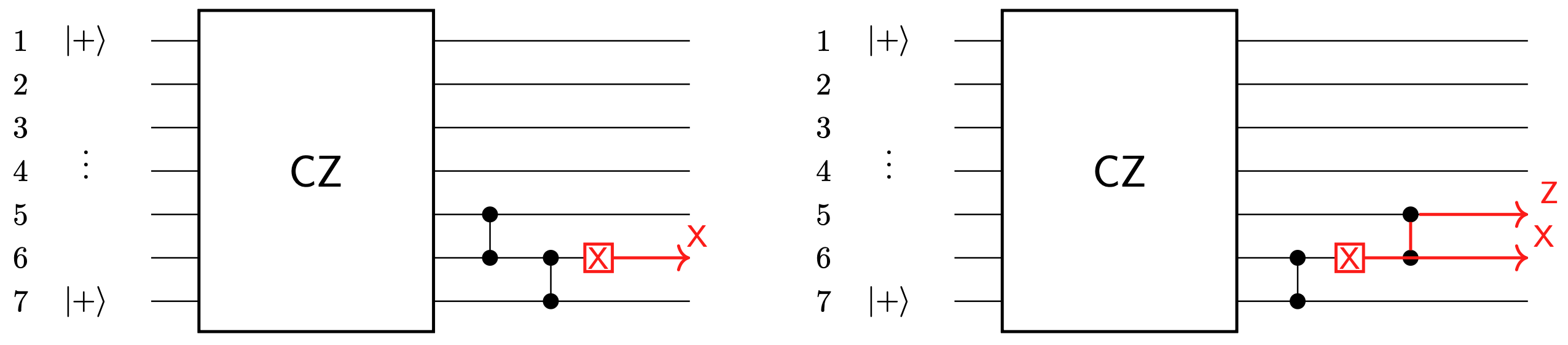}
\caption{Two gate orderings used to isolate depolarizing parameters associated with a central qubit. For an ordered triple $(5,6,7)$ with $(5,6),(6,7)\in E$, the relative order of $\CZ_{(5,6)}$ and $\CZ_{(6,7)}$ is swapped while keeping all other remaining gates fixed. Comparing the corresponding trap statistics yields equations that isolate the depolarizing strengths on the central qubit $6$ for the two incident edges.}
	\label{fig:orderings}
\end{figure}

\begin{figure}[H]
    \centering
    \includegraphics[scale=0.35]{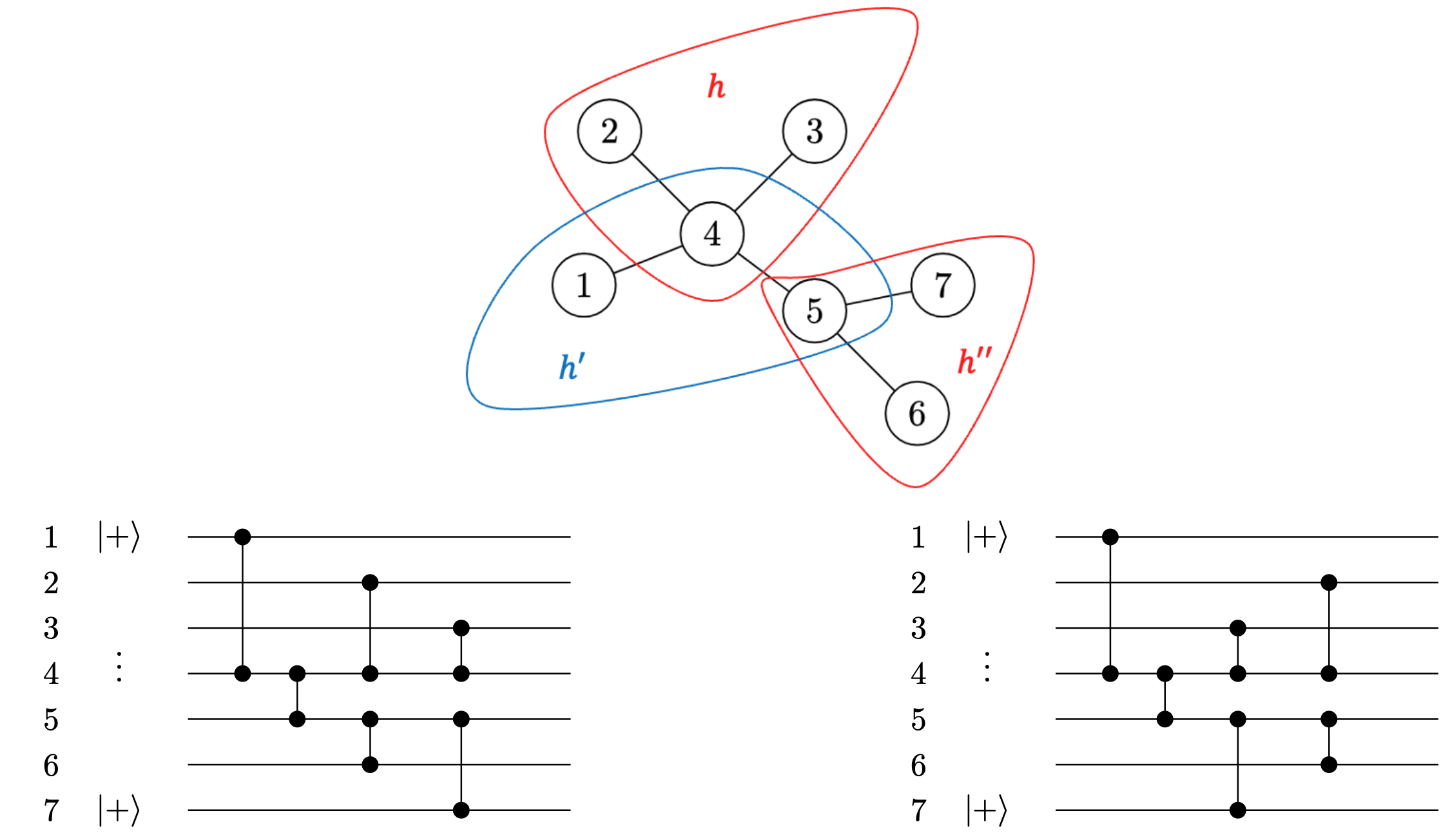}
    \caption{Parallelizing the noise parameter estimation. (Top) Construction of the graph \(H\) from the edges of \(G\). An ordered triple, e.g.~\(h\), covers two adjacent edges in \(G\). These triples form the vertices \(W\) of \(H\). There is an edge in \(H\) between \(h\) and \(h'\) as they share vertices in \(G\). On the contrary \(h\) and \(h''\) are not connected by an edge in \(H\). As a consequence it is possible to fix the order of the \(\CZ\) gates corresponding to the edges covered by \(h\) and \(h''\) independently, which implies parallelization. (Left) and (Right) show how, in such situation, two orderings are enough to extract \(\lambda_{(2,4),4}, \lambda_{(3,4),4}, \lambda_{(5,6),5},  \lambda_{(5,7),5}\).}
	\label{fig:single-color-ordering}
\end{figure}

Building on this remark, one can construct the graph \(H = (W,F)\) where vertices are elements of \(W\), and where \((h,h') \in F\) if \(h\) and \(h'\) have at least a vertex of \(G\) in common (see \cref{fig:orderings}). Then, twice the chromatic number of this graph is the number of orderings needed to recover all the depolarizing noise parameters. The reason is that within a color, 2 orderings are sufficient to extract the noise parameters for the central qubits (see \cref{fig:single-color-ordering}).

\begin{figure}[H]
    \centering
    \includegraphics[width=1\linewidth]{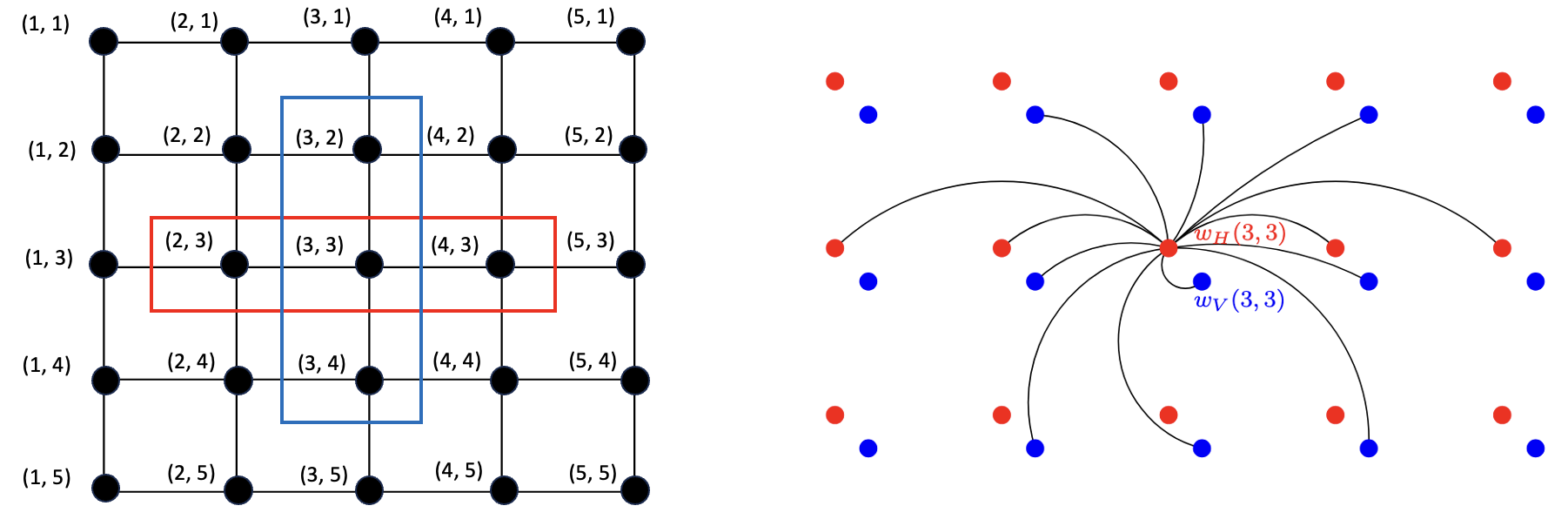}


    \caption{Possible construction of the graph $H$ for a 2D cluster-state graph $G$.
      (Top) The black vertices and edges are that of the 2D cluster-state. Qubits are indexed by their coordinates on the grid. The blue and red sets of qubits are the two triples that need to be considered for constructing the graph \(H\) from \(G\) at qubit \((3,3)\). Indeed, at qubit \((3,3)\) there are four \(\lambda\) parameters to estimate, corresponding to the four edges incident to this qubit. Here, each triple infers two different parameters. This being the case for all qubits, the vertices in \(H\) can be partitioned in two sets corresponding to horizontal and vertical triples. ()Bottom) The graph \(H\) constructed around the horizontal triple of qubit \((3,3)\) denoted \(w_H(3,3)\). The vertex \(w_V(3,3)\) corresponds to the vertical triple for qubit \((3,3)\). The other red and blue nodes correspond to the horizontal and vertical triples for the qubits of the graph \(G\). The neighborhood of \(w_H(3,3)\) is made of all triples that share a qubit with it. This amounts to \(w_H(1,3), w_H(2,3), w_H(4,3), w_H(5,3)\) for the horizontal ones and \(w_V(2,2), w_V(2,3), w_V(2,4), w_V(3,2), w_V(3,3), w_V(3,4), w_V(4,2), w_V(4,3), w_V(4,4)\). This shows that the degree of \(w_H(3,3)\) is 13. By symmetry, with periodic boundary conditions on \(G\), all nodes in \(H\) have degree 13. This means that the chromatic number is bounded by 14 and that 28 orderings are enough to gather all possible depolarizing parameters for the 2D cluster state. Only the edges incident to $w_H(3,3)$ have been pictured.}
    \label{fig:2d-cluster}
\end{figure}

A minimization over the possible graphs \(H\) of the chromatic number would then give the minimum number of orderings for finding all depolarizing parameters. Yet, as finding the chromatic number of generic graphs is \textsf{NP-Hard}, so is this minimization in the general case. One can nonetheless work this out for specific graphs. For the 2D cluster state, one can construct \(H\) and show that its maximum degree is 7, which implies that the chromatic number is at most 8, and at most 16 orderings are enough (See \cref{fig:2d-cluster} for a possible construction of \(H\)).

A similar construction would also yield a graphical representation of conditions to parallelize the parameter estimation for cross-talk noise.

\paragraph{Beyond depolarizing channel.}
In the case where the noise is not depolarizing but still acts on each qubit independently, the strategy above would not work. The reason is that pure \(\Z\) noise commutes with the \(\CZ\) gates and its effect on traps does not depend on the orderings. As a result, it is impossible to split the source of \(\Z\) noise between the various gates acting on a given qubit. One possible way to bypass this limitation is to realize that one could instead selectively replace a given \(\CZ_e\) by \(\CZ_e \circ \CZ_e \circ \CZ_e\). This would accumulate the effect of the \(\Z\) noise and offer a way to estimate it for this specific gate. Similar ``tricks'' could be played to possibly extract information about even more complex noise models as long as their noiseless combined effect is the identity.

\section{Numerical Analysis}

In this section, we numerically evaluate the proposed noise inference procedure on two representative examples. The first is a small, arbitrary graph that serves as a transparent validation of the proposed method. The second is a 2D cluster-state, used to assess the estimator's performance on a larger, more structured graph. Throughout this section, we work with an effective one-parameter per edge description, and denote by $\lambda_{e}$ the effective noise parameter associated with edge $e\in E$. This edge-level description is the effective parametrization used for the numerical reconstruction in this section, and serves as a compressed version of the more detailed notation introduced in Section~\ref{sec:Depnoise}. For a given gate ordering $C$ and trap qubit $v$, the corresponding trap bias is modeled as
\begin{equation}
	P_v(C)=\prod_{e\in S_v(C)} \lambda_e,
\end{equation}
where $S_v(C)$ is the set of edges whose propagated Pauli contributions affect the trap outcome at $v$ under ordering $C$. In the numerical simulations, the trap-failure probabilities are estimated from finite samples, and the corresponding empirical trap biases are used to reconstruct the parameters $\lambda_{e}$. The main purpose of this section is twofold. First, we verify that the ordering-based ratio estimator correctly isolates the contribution of individual edges. Second, we study how the reconstruction accuracy improves with the number of measurement samples.

\subsection{Arbitrary graph example}

\begin{figure}[H]
  \centering
  \includegraphics[scale=0.35]{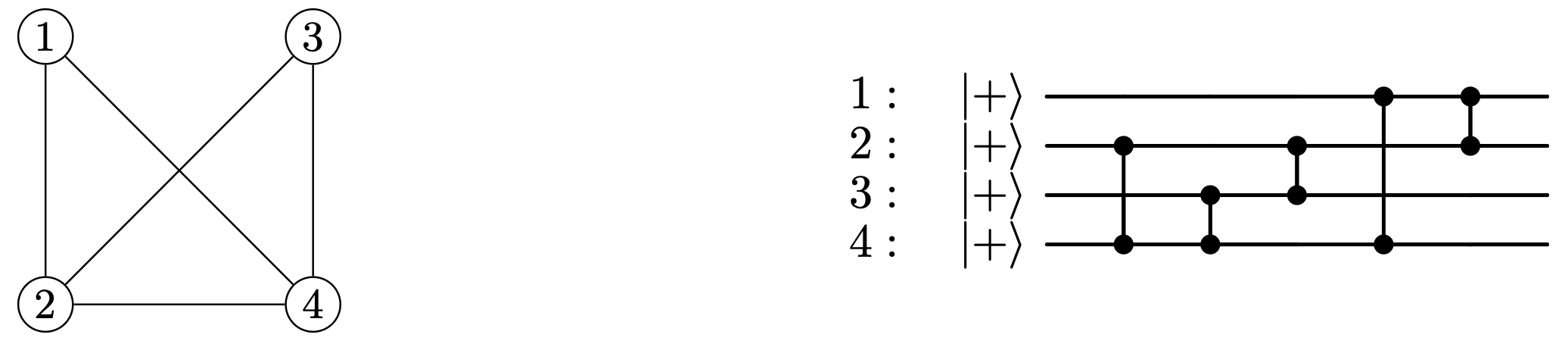}
  \caption{Graph state with edges \(E=\{(2,4),(2,3),(3,4),(1,4),(1,2)\)\} and all qubits initialized in \(\ket +\)}
  \label{fig:diamond-kite-direct-alternate}
\end{figure}

We first consider the graph shown in ~\cref{fig:diamond-kite-direct-alternate}, with vertex set $V=\{1,2,3,4\}$ and edge set $E=\{(1,2),(1,4),(2,3),(2,4),(3,4)\}$. The corresponding unknown effective noise parameters are $\lambda_{(1,2)},\ \lambda_{(1,4)},\ \lambda_{(2,3)},\ \lambda_{(2,4)},\ \lambda_{(3,4)}$.

For this example, we use a homogeneous one-qubit depolarizing noise model in order to make the reconstruction mechanism fully explicit. After each $\mathrm{CZ}$ gate, each involved qubit is subjected to a depolarizing channel with error probability $p_{\mathrm{err}}=0.002$. The associated effective parameter is then
$\lambda_{\mathrm{th}}=1-\frac{4}{3}p_{\mathrm{err}}=1-\frac{4}{3}(0.002)=0.997333\ldots$. Under this uniform setting, all edge parameters take the same value, $\lambda_{(1,2)}=\lambda_{(1,4)}=\lambda_{(2,3)}=\lambda_{(2,4)}=\lambda_{(3,4)}=\lambda_{\mathrm{th}}$.

We now illustrate how one of these parameters can be recovered from trap statistics. Consider the trap associated with qubit $1$, corresponding to the stabilizer generator $K_1=X_1\otimes Z_2\otimes Z_4$. For a reference ordering $C$, suppose that the propagated support contributing to the trap bias at qubit $1$ is $S_1(C)=\{(1,2),(1,4),(2,3),(2,4),(3,4)\}$, then $P_1(C)= \lambda_{(1,2)}\lambda_{(1,4)}\lambda_{(2,3)}\lambda_{(2,4)}\lambda_{(3,4)} \approx 0.9867376$.

Next, we choose another ordering $C'$ such that the contribution of edge $(2,3)$ is removed while all other contributions remain unchanged, namely $S_1(C')=S_1(C)\setminus\{(2,3)\}$. This gives
$P_1(C')=\lambda_{(1,2)}\lambda_{(1,4)}\lambda_{(2,4)}\lambda_{(3,4)}\approx 0.9893759$.
Taking the ratio of the two trap biases isolates the parameter of the missing edge:
\begin{equation}
	\widehat{\lambda}_{(2,3)}
	=
	\frac{P_1(C)}{P_1(C')}
	\approx
	\frac{0.9867376}{0.9893759}
	\approx 0.997333,
\end{equation}
which agrees with the theoretical value $\lambda_{\mathrm{th}}$.

The same procedure can be repeated for the other edges by selecting suitable trap qubits and pairs of orderings. This example, therefore, serves as a numerical validation of the estimator, i.e., the effective edge parameter is recovered directly from simulated trap statistics, and the reconstructed value matches the injected noise parameter.

\subsection{2D cluster-state example}

We now consider a 2D cluster-state with vertices arranged on a $W \times H$ square lattice as described in \cref{fig:2d-cluster} and edges connecting nearest neighbours horizontally and vertically. In this case, the graph is significantly larger, and the numerical study is used to assess the behavior of the reconstruction method under heterogeneous noise and finite sampling. As in the previous sections, multiple global orderings of the commuting $\mathrm{CZ}$ gates are generated from the ordered-triple construction and the associated conflict graph. These orderings are then used to produce distinct trap-bias equations from which the effective edge parameters are reconstructed.

In contrast to the arbitrary-graph example, the noise is now non-uniform. Specifically, after each $\mathrm{CZ}$ gate, the one-qubit depolarizing error probability acting on each involved qubit is sampled independently from a Gaussian distribution, 
\begin{equation}\label{equ:gaus}
p \sim \mathcal{N}\!\left(10^{-2},\,2\times 10^{-3}\right), \mathcal{N}\!\left(10^{-2},\,2\times 10^{-4}\right), \dots, \mathcal{N}\!\left(10^{-2},\,2\times 10^{-7}\right).
\end{equation}
In the numerical implementation, the sample values are truncated to the physical interval $[0,1]$ whenever necessary. For each sampled value of $p$, the corresponding one-qubit depolarizing factor is computed as $\lambda = 1-\frac{4}{3}p$.

Thus, the simulation is generated from the underlying qubit after gate noise samples rather than from a single uniform noise parameter. The quantities $\lambda_{e}$ appearing in the reconstruction are therefore effective edge-level parameters inferred from the resulting trap-bias equations. For a given ordering $C$ and trap qubit $v$, the trap-failure probability is estimated from the sampled outcomes as
\begin{equation}
	\widehat{p}_{\mathrm{fail}}(v,C)
	=
	\frac{N_{\mathrm{fail}}(v,C)}{N_{\mathrm{shots}}(v,C)},
\end{equation}
and the corresponding empirical trap bias is $\widehat{P}_v(C)=1-2\widehat{p}_{\mathrm{fail}}(v,C)$.
Using the family of generated orderings, these empirical trap biases produce an overdetermined system for the unknown parameters $\lambda_{e}$, whose solution yields the reconstructed values $\widehat{\lambda}_e$.

\begin{figure}[!h]
	\centering
    \begin{subfigure}{0.48\textwidth}
	\includegraphics[width=\linewidth]{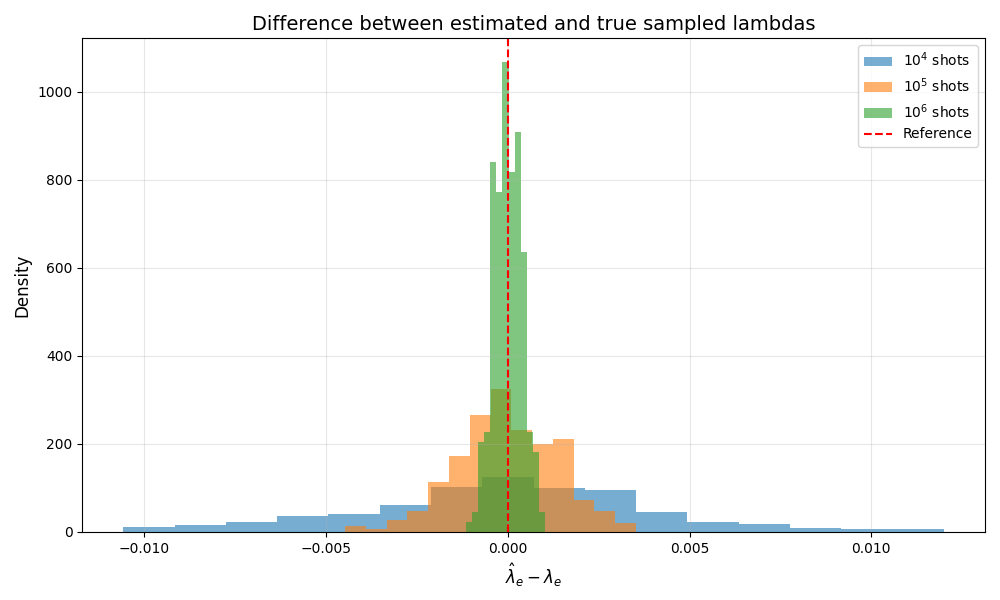}
    \end{subfigure}
\hfill
\begin{subfigure}{0.48\textwidth}
	\includegraphics[width=\linewidth]{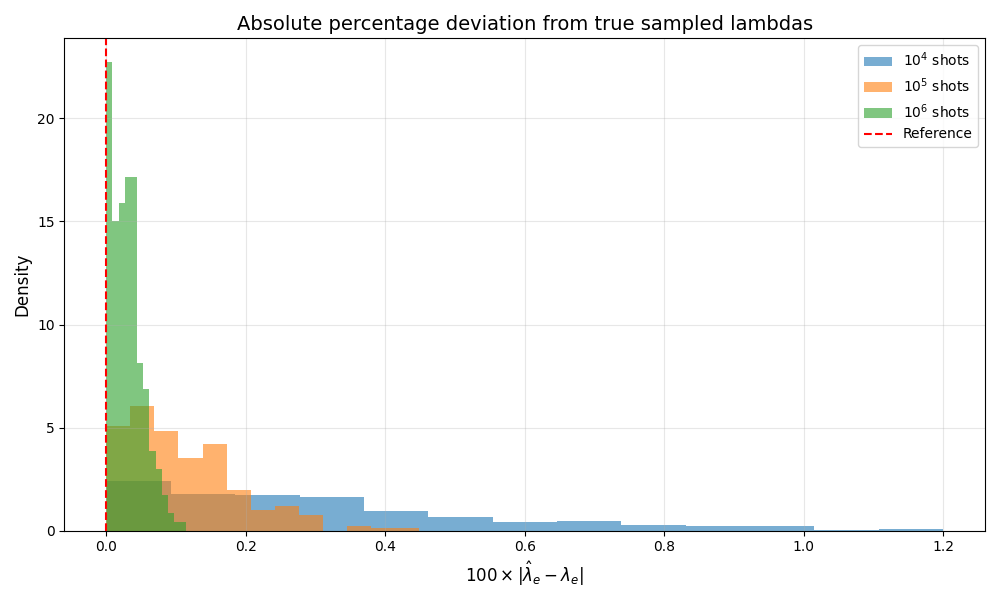}
\end{subfigure}
\caption{Distribution of the reconstruction differences $\widehat{\lambda}_e-\lambda_e$
(left) and of the absolute deviations $100\times |\widehat{\lambda}_e-\lambda_e|$ (right)
for the $12\times 12$ 2D cluster-state, using $10^4$, $10^5$, and $10^6$ shots, when the
post-$\CZ$ one-qubit depolarizing error probabilities are sampled independently from
$\mathcal{N}(10^{-2},\,2\times 10^{-3})$. The dashed red line marks zero difference. As the
number of shots increases, the signed-error distribution sharpens around zero and the absolute
deviations become more concentrated near zero, indicating improved reconstruction accuracy.}
\label{fig:lambda_diff_by_shots}
\end{figure}

\begin{figure}[!h]
	\centering
    \begin{subfigure}{0.48\textwidth}
	\includegraphics[width=\linewidth]{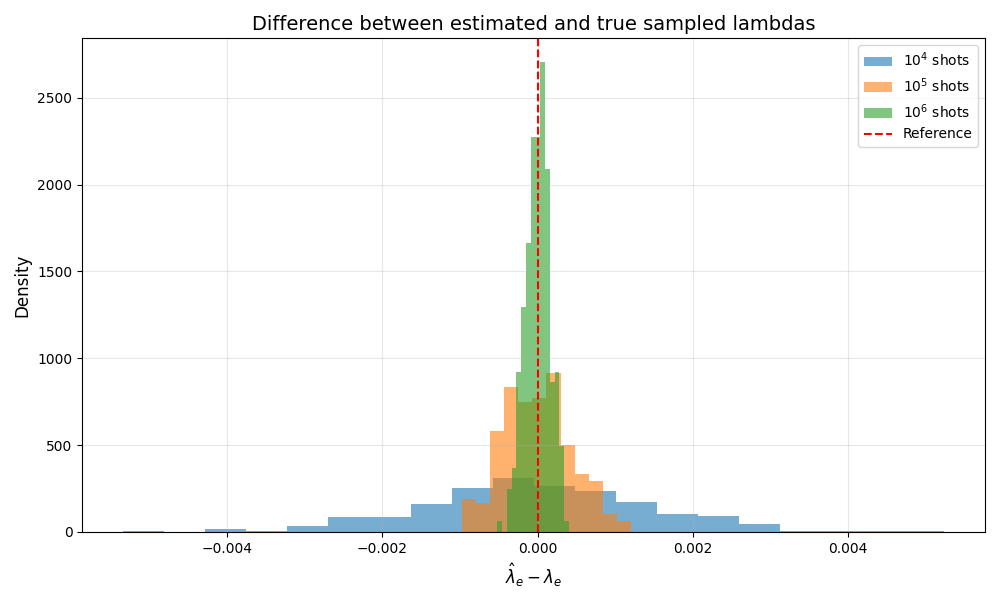}
    \end{subfigure}
\hfill
\begin{subfigure}{0.48\textwidth}
	\includegraphics[width=\linewidth]{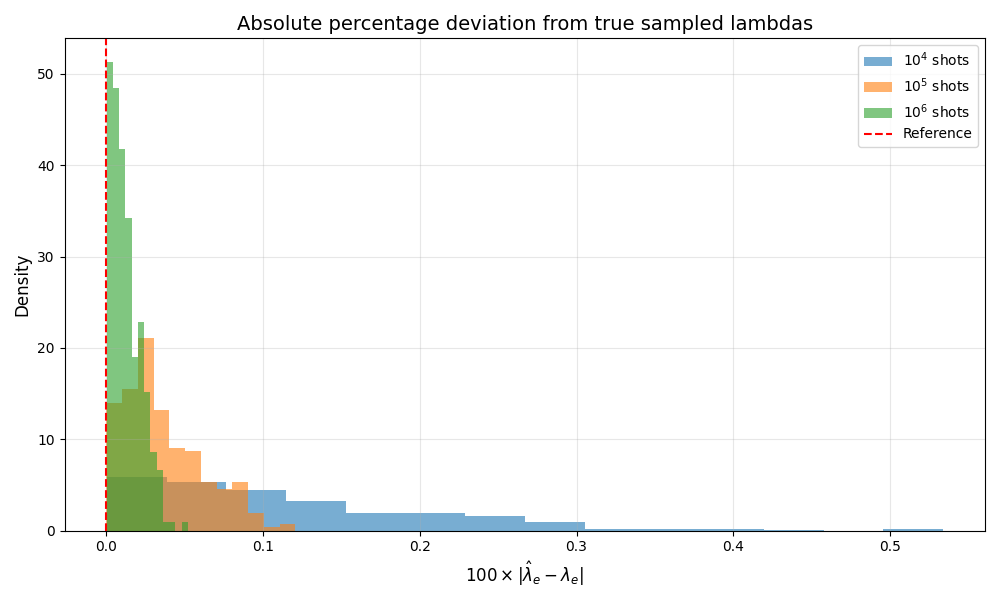}
\end{subfigure}
\caption{Distribution of the reconstruction differences $\widehat{\lambda}_e-\lambda_e$
(left) and of the absolute deviations $100\times |\widehat{\lambda}_e-\lambda_e|$ (right)
for the $12\times 12$ 2D cluster-state, using $10^4$, $10^5$, and $10^6$ shots, when the
post-$\CZ$ one-qubit depolarizing error probabilities are sampled independently from
$\mathcal{N}(10^{-2},\,2\times 10^{-7})$. The dashed red line marks zero difference. This is
the most concentrated noise ensemble among the five cases, and the reconstruction errors shrink
further as the shot count increases.}
\label{fig:lambda_abs_diff_by_shots}
\end{figure}

For the numerical study, we take a cluster-state of width $W=12$ and depth $H=12$.
Reconstruction quality is evaluated from the distributions of the signed differences
$\widehat{\lambda}_e-\lambda_e$ and the absolute deviations
$100\times |\widehat{\lambda}_e-\lambda_e|$ over all reconstructed edges.
For each of the five Gaussian noise spreads in Eq.~\eqref{equ:gaus}, we compare
three shot budgets: $10^4$, $10^5$, and $10^6$. Figs.~\ref{fig:lambda_diff_by_shots}~$-$~\ref{fig:lambda_abs_diff_by_shots} report the corresponding
histograms.

Several features are consistent across all five noise ensembles. First, the histograms of
$\widehat{\lambda}_e-\lambda_e$ remain centered close to zero, indicating that the estimator
does not exhibit a visible systematic bias. Second, the distributions narrow markedly as the
number of shots increases from $10^4$ to $10^6$, which is the expected finite-sampling
behavior. The same trend is visible in the histograms of the absolute deviations, whose mass
becomes progressively more concentrated near zero as the shot count increases. This shows that
the reconstructed parameters become more stable and more accurate when more trap rounds are
available. 

A second trend appears when comparing Figs.~\ref{fig:lambda_diff_by_shots}~$-$~\ref{fig:lambda_abs_diff_by_shots} with one another. As the variance of the
underlying Gaussian sampling is reduced from $2\times 10^{-3}$ to $2\times 10^{-7}$, the
reconstructed distributions become increasingly concentrated. This is consistent with the fact
that smaller spread in the injected qubit-level noise yields a more homogeneous effective
edge-level model, which is easier to reconstruct from the available trap statistics. Taken
together, these results support the effectiveness of the proposed inference method on a large
and structured graph. In other words, the ordering-based reconstruction remains stable under heterogeneous
noise and improves predictably with increasing sample size.

\begin{figure}[!h]
    \centering
    \begin{subfigure}{0.48\textwidth}
    \centering
	\includegraphics[width=\linewidth]{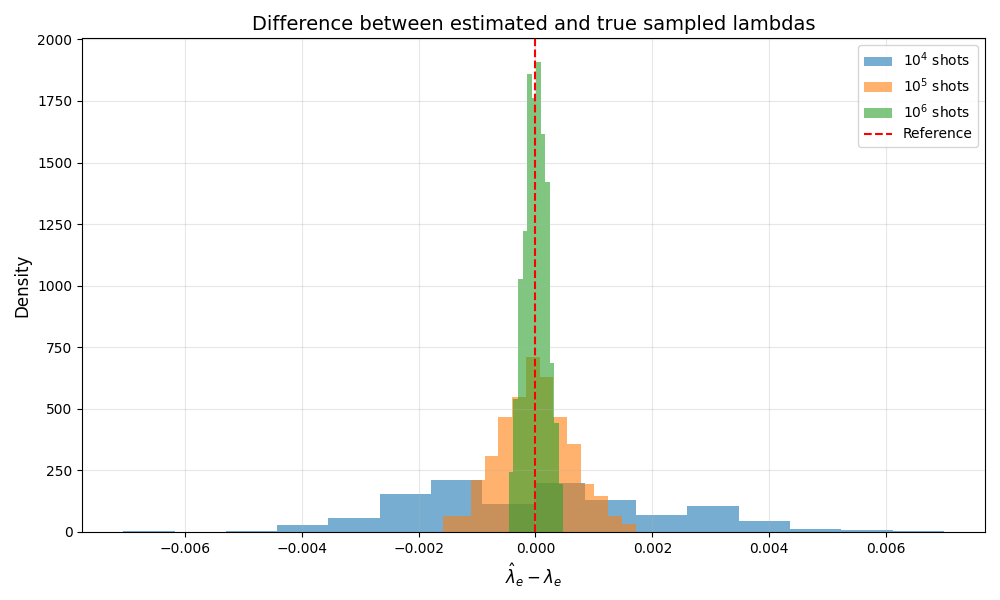}
    \end{subfigure}
\hfill
\begin{subfigure}{0.48\textwidth}
	\centering
	\includegraphics[width=\linewidth]{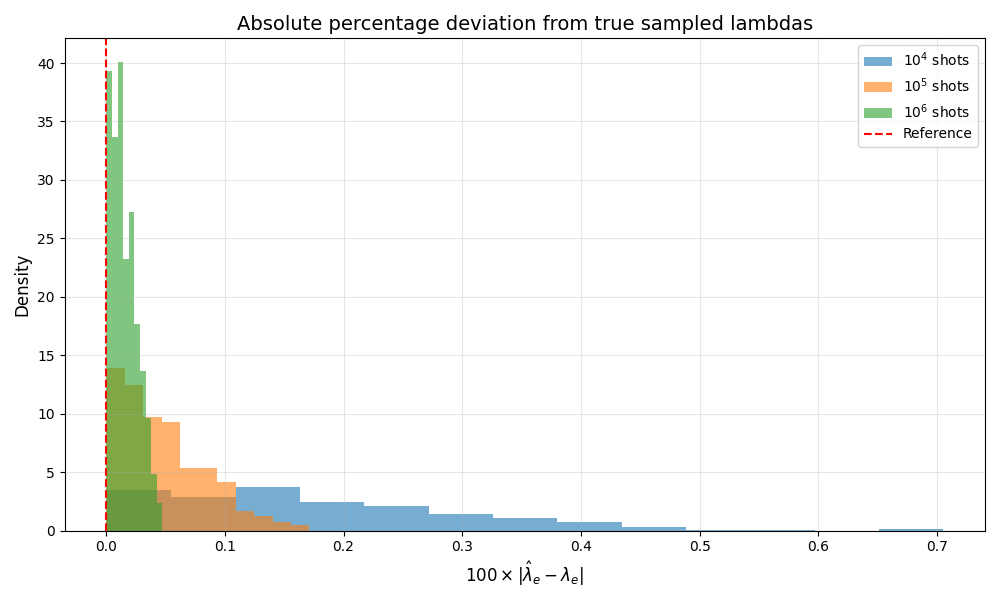}
\end{subfigure}
\caption{Distribution of the reconstruction differences $\widehat{\lambda}_e-\lambda_e$
(left) and of the absolute deviations $100\times |\widehat{\lambda}_e-\lambda_e|$ (right)
for the $12\times 12$ 2D cluster-state, using $10^4$, $10^5$, and $10^6$ shots, when the
post-$\CZ$ one-qubit depolarizing error probabilities are sampled independently from
$\mathcal{N}(10^{-2},\,2\times 10^{-4})$. The dashed red line marks zero difference. The
reconstruction is centered near zero and becomes progressively more concentrated as the shot
count increases.}
\end{figure}
    \begin{figure}[!h]
	\centering
    \begin{subfigure}{0.48\textwidth}
	\includegraphics[width=\linewidth]{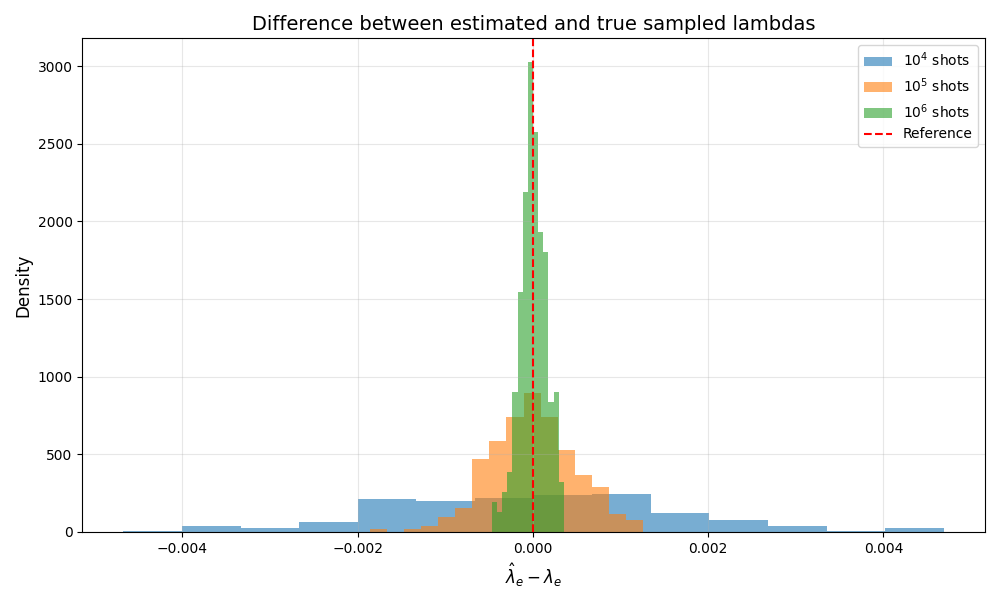}
    \end{subfigure}
\hfill
\begin{subfigure}{0.48\textwidth}
	\includegraphics[width=\linewidth]{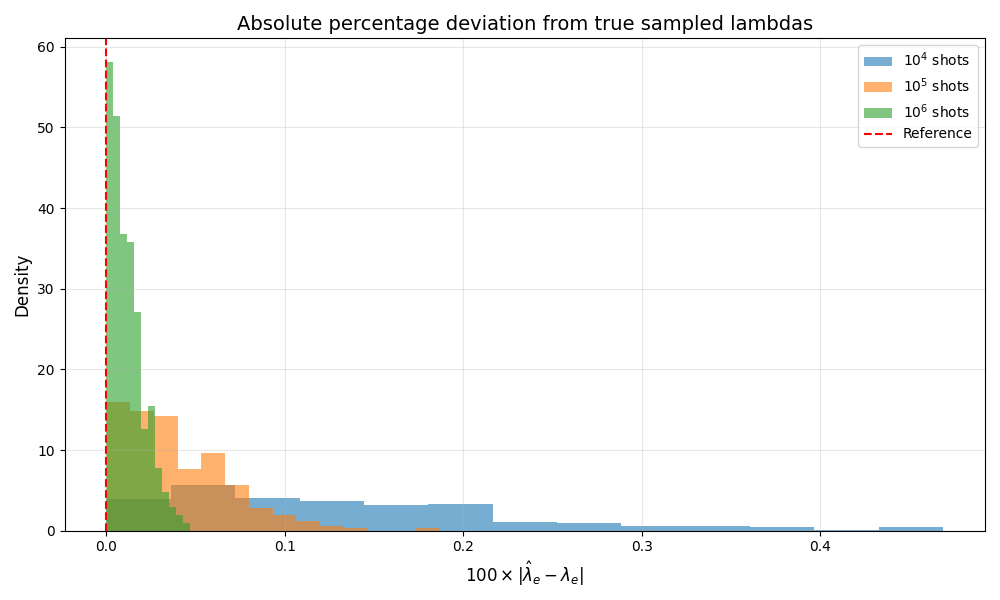}
\end{subfigure}
\caption{Distribution of the reconstruction differences $\widehat{\lambda}_e-\lambda_e$
(left) and of the absolute deviations $100\times |\widehat{\lambda}_e-\lambda_e|$ (right)
for the $12\times 12$ 2D cluster-state, using $10^4$, $10^5$, and $10^6$ shots, when the
post-$\CZ$ one-qubit depolarizing error probabilities are sampled independently from
$\mathcal{N}(10^{-2},\,2\times 10^{-5})$. The dashed red line marks zero difference. The
signed and absolute reconstruction errors both contract as more samples are collected.}
\end{figure}
\begin{figure}[!h]
	\centering
    \begin{subfigure}{0.48\textwidth}
	\includegraphics[width=\linewidth]{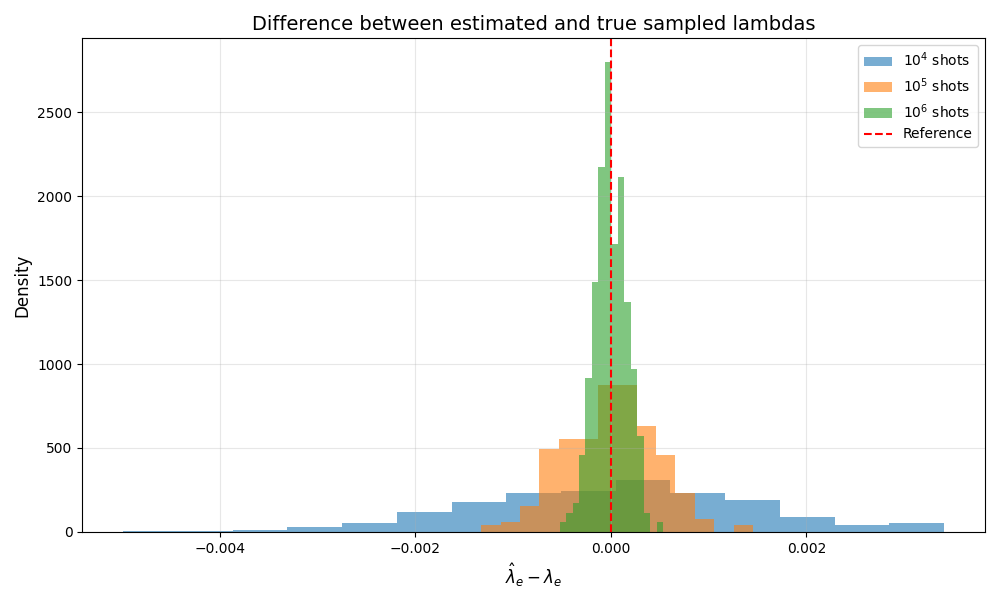}
    \end{subfigure}
\hfill
\begin{subfigure}{0.48\textwidth}
	\includegraphics[width=\linewidth]{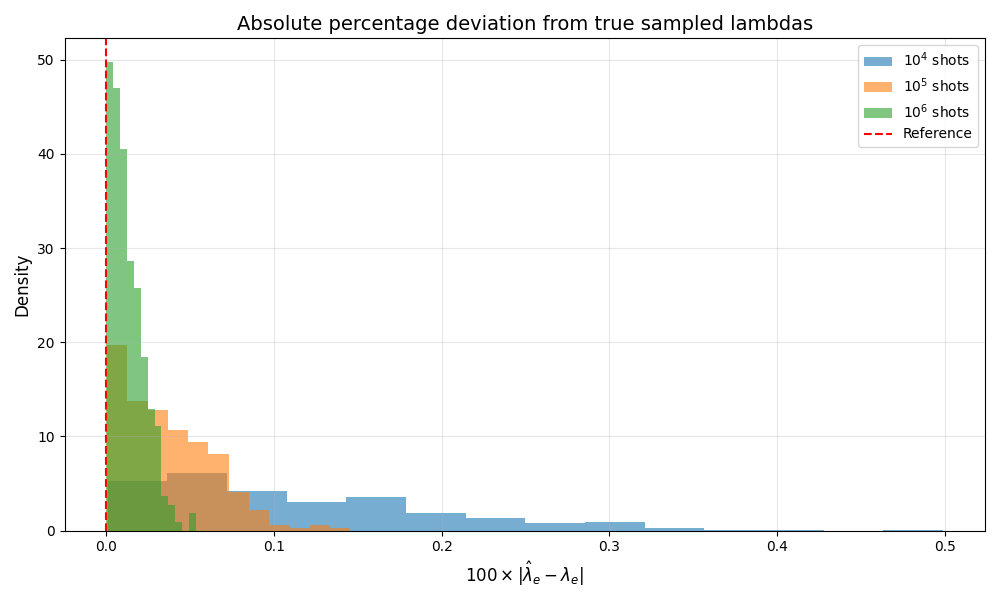}
\end{subfigure}
\caption{Distribution of the reconstruction differences $\widehat{\lambda}_e-\lambda_e$
(left) and of the absolute deviations $100\times |\widehat{\lambda}_e-\lambda_e|$ (right)
for the $12\times 12$ 2D cluster-state, using $10^4$, $10^5$, and $10^6$ shots, when the
post-$\CZ$ one-qubit depolarizing error probabilities are sampled independently from
$\mathcal{N}(10^{-2},\,2\times 10^{-6})$. The dashed red line marks zero difference. The
histograms remain centered near zero, while increasing the number of shots reduces the spread
of the reconstruction error.}
\end{figure}

\section{Discussion}
This work shows that the perceived overhead of verification, as measured by the increase in rounds to obtain a verified outcome, is not actually a real overhead, as it can serve other purposes, such as partial recalibration of the device. This is a significant advantage for the service provider as it has the potential to greatly reduce the downtime of the machine. More fundamentally, it shows that the information that is gathered about the behavior of the machine by verification protocols is highly dependent on what one is willing to assume. The client does only obtain an accept or reject bit, while the server can update a full error model with it. This indeed further strengthens the link between verification and error detection put forth in~\cite{KKLM22unifying}.

This work opens new future directions, such as proof-of-concept implementation of \cref{proto:vbqc-aces} on real devices and the obtaining of effective certified noise maps that could be used to perform verified error mitigation observable estimation. It also naturally asks whether other noise estimation techniques can be integrated into verification protocols, whether they are parametric or also allow the discovery of the noise structure itself.

On the more practical side, several optimizations and limitations should be taken into account to ensure the number of live qubits remains manageable in spite of the necessity of performing entangling gates in different orders, some orderings are clearly impractical if they lead to keeping some qubits live for a very long time, or if the number of live qubits exceeds the capacity of the machine itself.  We leave this open for future work.


\paragraph{Acknowledgements.} We thank Maxime Garnier and Thierry Martinez for the fruitful discussion on numerical analysis. This work was funded in part by the European Union under Grant No. 101080142 (EQUALITY) and by France 2030 under the French National Research Agency award number ANR-22-PNCQ-0002 (HQI).

\bibliographystyle{alpha}
\bibliography{qubib}
\appendix

\section{Robust Verification of Quantum Computations}
\label{sec:ubqc}

\paragraph{Measurement Based Quantum Computation.}
An MBQC algorithm, also called \emph{measurement pattern}, consists of a graph $G = (V,E)$, two vertex sets $I$ and $O$ defining input and output vertices, a list of angles $\{\phi_v \}_{v \in V}$ with $\phi_v \in \Theta := \{k\pi/4\}_{0 \leq k \leq 7}$ and a function \(f:O^c\rightarrow I^c\) called \emph{flow}.

\begin{figure}[!h]
    \centering
    \includegraphics[width=0.5\linewidth]{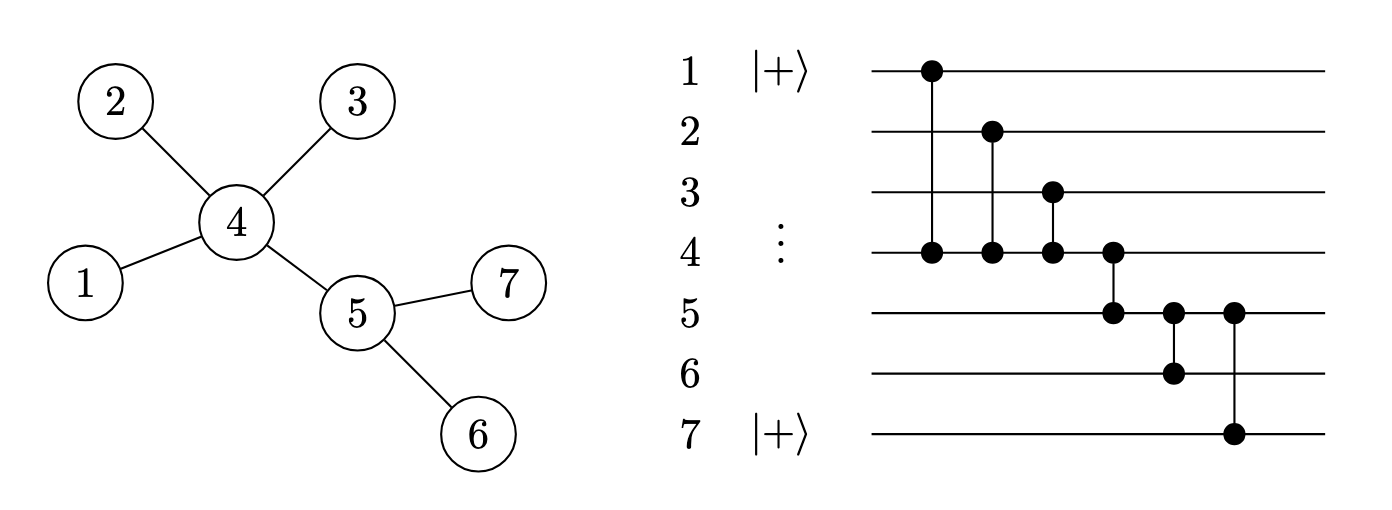}
    \caption{Translation of a graph state (left) into a circuit (right). Vertices correspond to qubits prepared in $\ket{+}$, and CZ gates between qubits correspond to vertices that share an edge in the graph.}
    \label{fig:graph}
\end{figure}

To run it, the client instructs the server to prepare the graph state $\ket{G}$: for each vertex in $V$, the server creates a qubit in the state $\ket{+}$ and performs a $\CZ$ gate for each pair of qubits in $E$. A canonical example of such a resource state is shown in Figure \ref{fig:graph}. The client then asks the server to measure each qubit of $V$ along the basis $\left\{\ket{+_{\phi'_v}}, \ket{-_{\phi'_v}}\right\}$ in the order defined by the flow of the computation, with $\ket{+_\alpha} = (\ket 0 + e^{i\alpha}\ket 1)/\sqrt 2$. The corrected angle $\phi'_v$ is given by $\phi'_v = (-1)^{s_v^X}\phi_v + s_v^Z\pi$ for binary values of $s_v^X$ and $s_v^Z$ that depend only on the outcomes of previously measured qubits and the flow---the interested reader is referred to~\cite{HEB04multiparty,DKP07measurement-calculus} for details about the flow.

As shown in \cite{DKP07measurement-calculus}, the MBQC model is equivalent to the circuit model, so that any \(\mathsf{BQP}\) algorithm in the circuit model can be translated in the MBQC model with at most polynomial overhead.

\paragraph{Hiding the Computation.}
In MBQC, a computation can be easily hidden if, instead of the server preparing each qubit, the client (i) for all $v \in V$ sends $\ket{+_{\theta_v}}$ with $\theta_v$ chosen uniformly at random in $\Theta$; (ii) asks the server to measure the qubits in the basis defined by the angle $\delta_v = \phi'_v + \theta_v + r_v\pi$ for $r_v$ a random bit, while keeping $\theta_v$ and $r_v$ hidden from the server; and (iii) uses $s_v = b_v \oplus r_v$ where $b_v$ is the measurement outcome to compute $s_v^X$ and $s_v^Z$ defined above. The random sampling of $\theta_v$ implements a One-Time Pad for $\phi'_v$ while $r_v$ does the same for the measurement outcomes. 

Indeed, this initial idea can be turned into a protocol that provides blindness even when the initial unencrypted qubits are not in the \(\ket +\) state. This is the Universal Blind Quantum Computing protocol introduced in~\cite{BFK09universal}.  It works by adding a random bit flip \(\X\) per qubit, updating the measured angles accordingly, and also one-time-padding the returned output.

\begin{protocol}[Universal Blind Quantum Computing (UBQC)]
  \label{proto:ubqc}
\item
  \begin{algorithmic}[0]
    \State \textbf{Public information}: A graph \(G\) and a flow \(f\).
    \State \textbf{Client's input}: \(\rho\) the initial state of the qubits at nodes of \(G\), a measurement pattern using the flow \(f\), that is a list of measurement angles \(\{\phi_{v}\}_{v\in V}\).

    \Procedure{State Encryption}{}
    \State For each \(v\in V\) sample the following encryption secrets  \(a_v \sample \{0,1\}\), \(r_v \sample \{0,1\}\), \(\theta_v \sample \Theta\).
    \State Apply \(\Z(\theta_{v}) \X^{a_{v}}\) to qubit at \(v\).
    \EndProcedure

    \Procedure{Graph State Creation}{}
    \State The server applies \(\CZ\) gates for all edges of $G$.
    \EndProcedure
    
    \Procedure{Encrypted Computation Orchestration}{}
    \State In the order of the flow, the client sends a measurement angle for qubit \(v\) as \(\delta_{v} = \phi_v' + \theta_v + r_v \pi\), and receives \(b_v\) from the server, corresponding to the encoded measurement outcome for this measurement. Here, \(\phi_v'\) is computed as above using \(s_u = b_u \oplus r_u\) as the decrypted measurement outcome for the measurement occurring at \(u\).
    \EndProcedure
  \end{algorithmic}
\end{protocol}

\paragraph{Generating Test Rounds}
\label{sec:test-rounds}
Given a graph \(G=(V,E)\), and a vertex \(v \in V\) such that \(v\) is prepared in \(\ket +\) and all its neighbors in \(G\) are prepared in \(\ket 0\), it is easy to verify that after the creation of the graph all these qubits remain unchanged. This is because a \(\CZ\) gate applied to a qubit in \(\ket 0\) leaves both untouched. If \(\X\) is then measured on \(v\), this would yield the +1 outcome. We call such preparation of \(v\) and neighbors a \emph{trap}, and the outcome of the \(\X\) measurement on \(v\) the \emph{trap outcome}; \emph{passed} if +1 and \emph{failed} if -1. 

Several traps can be inserted in a given round as long as they are not neighbors. Such a round is then called a \emph{test} round. For rVBQC to provide security, each vertex of \(G\) must be a trap in a constant fraction of the test rounds, as all traps cannot be placed in a single round. Indeed, for a \(k\)-colorable graph, \(k\) different types of test rounds are enough to fulfill this constraint. This gives the following routine:
\begin{routine}[Test Round Sampling]
  \label{routine:test-sampling}
  \item
  \begin{algorithmic}[0]
    \State A \(k\)-colorable graph \(G\) and a \(k\)-coloring \(\{V_i\}_i, \ \dot{\cup}_i V_i = V\)
    \Procedure{Test Round Sampling}{}
    \State \(i \sample [k]\)
    \State Output the preparation of traps at \(v \in V_i\)
    \State Output a measurement pattern so that all \(v \in V_i\) are measured with angle 0 corresponding to measuring \(\X\), setting the rest to random values in \(\Theta\).
    \EndProcedure
  \end{algorithmic}
\end{routine}

\end{document}